\documentclass[aps,pre,showpacs,floatfix,twocolumn,superscriptaddress]{revtex4-2}
\usepackage{mathrsfs}

\usepackage{graphicx}
\usepackage{bm,amsmath,amssymb}
\usepackage{dcolumn}
\usepackage{color}
\usepackage{bm}

\begin{document}
\title{Phase transitions in 3D Ising model with cluster weight by Monte Carlo method}
\author{ Ziyang Wang}
\affiliation{
Key Laboratory of Quantum Information, University of Science and Technology of China, Chinese Academy of Sciences, Hefei 230026, PR China}
\affiliation{College of Physics and Optoelectronics, Taiyuan University of Technology, Shanxi 030024, China}

\author{ Le Feng}
\affiliation{College of Physics and Optoelectronics, Taiyuan University of Technology, Shanxi 030024, China}

\author{ Wanzhou Zhang }
\thanks{zhangwanzhou@tyut.edu.cn}

\affiliation{
Key Laboratory of Quantum Information, University of Science and Technology of China, Chinese Academy of Sciences, Hefei 230026, PR China}
\affiliation{College of Physics and Optoelectronics, Taiyuan University of Technology, Shanxi 030024, China} 

\author{Chengxiang Ding}
\thanks{dingcx@ahut.edu.cn}
\affiliation{
School of Science and Engineering of Mathematics and Physics, Anhui University of Technology, Maanshan 243002, China}

\date{\today}
\begin{abstract}
A cluster weight Ising  model is proposed by introducing an additional cluster weight in the partition function
of the traditional Ising model. It is equivalent to the O($n$) loop model or $n$-component face cubic loop model on the two-dimensional lattice,
but on the three-dimensional lattice, it is still not very clear whether or not  these models have the same universality.
In order to simulate the cluster weight Ising model and search for new universality class,  we  apply a cluster algorithm, by combining  the color-assignation and the Swendsen-Wang methods. The dynamical exponent for the absolute magnetization is estimated to be  $z=0.45(3)$ at $n=1.5$, consistent with that by  the traditional Swendsen-Wang methods. 
The numerical estimation of the thermal exponent $y_t$ and magnetic exponent $y_m$,
show that the universalities of the  two models on the three dimensional lattice are 
  different.
 We obtain the global phase diagram containing paramagnetic  and ferromagnetic phases. The phase transition between the two phases are second order at $1\leq n< n_c$ and first order at $n\geq n_c$, where $n_c\approx 2$.  The scaling dimension $y_t$ equals to the system dimension $d$  when the first order transition occurs.  Our results are helpful in the understanding of some traditional statistical mechanics models. 
\end{abstract}
\pacs{05.50.+q, 64.60.Cn, 64.60.De, 75.10.Hk}
\maketitle
\section{Introduction}
\label{intro}
A basic task in statistical physics is revealing the universalities of a many theoretical models describing the common properties of
different kinds of materials. The most original and standard model in statistical physics is the Ising model, proposed by Ising in the year
 1925\cite{ising}.
The model was generalized to a large variety of models, such as the O($n$) spin model initially defined  by Stanley \cite{on}  as $n$-component  spins 
 interacting  in an isotropic way. Another interesting model is the $n$-component face cubic model, which is usually defined as a Hamiltonian containing two nearest-neighbor interactions
between $n$-component spins  that point to the faces
of an $n$-dimensional hypercube  \cite{ncu2,ncubic}. Face cubic model's counterpart model is a corner cubic model with spins pointing to the corners instead of faces 
of the hypercube    \cite{ncubic,cornerc}.

The critical properties of the O($n$) spin   model and $n$-component face cubic model have been studied and compared extensively in the language of  graph  by expanding the partition function in power and integrating the spin variables.  The O($n$) spin model should be able to  be mapped to loop model~\cite{onsq,guosquare lattice,Zhe Fu}, where the parameter $n$ is not restricted to integers.
Similar mapping exists from $n$-component face cubic model to  the so called cubic-loop model named by Ref.~\cite{cubicguowenan}  or  Eulerian bond-cubic model~\cite{ding,ding13}, as each vertex (site) connects even number of bonds.
On the square lattice, in the range $1 \leq n < 2$, the O($n$) loop model and $n$-component face cubic model~\cite{ding} belong to the same universality class, and the critical exponents are expected  to be obtained by
mapping the model to Coulomb gas model ~\cite{nienhuis}. The difference of the two models start at $n=2$ because the O(2) spin model undergoes
a Berezinskii-Kosterlitz-Thouless(BKT) transition ~\cite{kt1,kt2,kt3} while
$n$-component face cubic model undergoes a second-order transition ~\cite{ding}.
For $n>2$ on the square lattice, there is no physical phase transition for O($n$) loop model ~\cite{dingd,Wenan Guo}, but the  face cubic loop model undergoes a first-order transition ~\cite{temperature exponent}.
In three-dimensions, O($n$) symmetry can lead to continuous transitions at very large $n$ ~\cite{On3d,Emergent}, while the cubic symmetry makes the transition
discontinuous when $n>{n_c}$ with $n_c\approx2.89$ ~\cite{ncomponent}.

In order to access the rich critical properties of loop model,  the local updates of  Monte Carlo simulation  is performed although it is a difficult and interesting task due to  the non-local weight
in the partition sum of the model~\cite{dingd}. To solve this problem, a cluster algorithm  combining the tricks of Swendsen-Wang algorithm ~\cite{SW} and
`coloring method' is proposed  by Deng et al ~\cite{cluster simulation}. In this algorithm, the microstate (configuration) of the loop model is represented
by the configuration of Ising spins, the loops are regarded as the domain walls~\cite{domain walls,Geometric properties} of the Ising clusters. However, a problems arises naturally, such representation is only
applicable in two dimensional  honeycomb lattices  because there is no loop intersection phenomenon; in three dimensions with  maximum  coordination number 3, due to the special topology, the  loop model was simulated in a way of resorting to other methods, for instance, the worm algorithm~\cite{YaDong Xu,worm monte}.

In this paper, we pay special attention to the Ising representation. In fact, the two dimensional loop model 
can be regarded as an Ising model with cluster weight, which we will explain in detail in the next section.
Generalization of such a `cluster weighted Ising' (CWI) model to three dimension is applicable and straight forward,
and the cluster algorithm is still applicable for it.
We will investigate this model by Monte Carlo simulations,
and compare its critical properties with the results from the loop models~\cite{ding, On3d}.

The outline of this work is as follows. Section~\ref{sec:Ising model with cluster weight} introduces
the  loop model and the Ising representation on the honeycomb and cubic lattices, the  proposed CWI model.  The difficulty of simulating the loop model in non-planar graphs is also described. 
Section~\ref{sec:algorithm} describes the cluster-update algorithm  and 
 several sampled observables in our Monte Carlo simulations. Numerical results
are then presented in Sec.~\ref{sec:results}. The global phase diagram are shown  and the  critical exponents for the first-order and second-order transitions are  presented. The efficiency of the algorithm  and how to get the error bars are also discussed.  Conclusive comments are made in Sec.~\ref{sec:conclusion}.

\section{Ising model with a cluster weight}
\label{sec:Ising model with cluster weight}
\begin{figure}[htpb]
 \centering

\hskip -1cm
\includegraphics[scale=0.35]{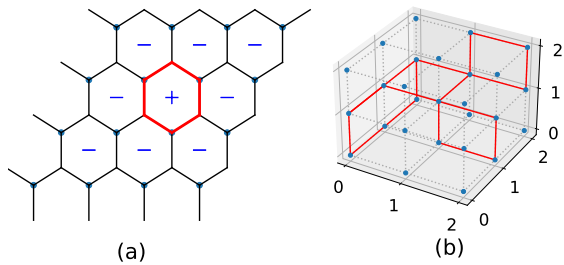}
\caption{(Color online) 
 (a) On a honeycomb lattice, a loop  denoted by red bonds forms  with the number of clusters $n_c=2$ and and the number of loops $l=1$ (b) 
    For the cubic lattices, the graph may not satisfy the the  planarity, required by the relationship $n_c= l + 1$.}
 
 \label{snap}
 \end{figure}
 
Starting at the partition function of the    loop model, 
\begin{equation}
Z_{\text{loop}}=\sum_G x^bn^l, \label{loop}
\end{equation}
where $l$ is the number of loops.
The loop configuration on the honeycomb lattice can be represented by the Ising configuration on the
triangular lattice (dual lattice of the honeycomb lattice), and the loops are just the domain walls~\cite{domain walls,Geometric properties} of the Ising clusters. The number of the Ising clusters $n_c$
is just the number of loops $l$ plus 1, namely $n_c=l+1$.

For the honeycomb lattices,  intersecting
loops does not emerge as shown in Fig.~\ref{snap} (a). 
The reason is that each configuration 
consists Eulerian-bond graph, ``Eulerian`` means each site (vertex) is connected to  even number of bonds.

However, for the square lattice, similar to the cubic lattice show in  Fig.~\ref{snap} (b), 
loop intersecting  will occur. This will cause difficulties in  counting the number of 
loop during simulations. Not only square lattice, 
any lattices with degree more than 3 will cause 
such confusion. Namely, each site connects more than 3 sites. 

One way to solve such a question 
is not using the term ``number of loops" in  Eq (1)  because such a quantity is not well defined on graphs with degree above 3.  The correct term is ``cyclomatic number", defined to be the minimum number of edges required to be deleted from a graph in order to obtain a forest~\cite{cnumber}.  The value of ``cyclomatic number'' is 
$l=e-n+1$, where $e$ and $n$ are the numbers of bonds and the sites in the lattices, respectively.
On graphs of maximum degree 3, the cyclomatic number does indeed count the number of loops $l$.

On the other hand, even using the definition of ``cyclomatic number'' $l$, the  relation $n_c = l +
1$ still holds   for the square and honeycomb
lattices, but  this is only true for
planar graphs.  The requirement of planarity is very crucial.
For the  cubic lattices, the graph may not satisfy the requirement. 

Here, we propose to performing direct research 
in the language of Ising clusters in the dual lattices rather than loop language.
The   partition function of  the CWI model proposed reads,
\begin{equation}
Z_{\text{CWI}}=\sum_{\{S_i\}} \exp(-H)n^{n_c},\label{cwi}
\end{equation}
with the reduced Hamiltonian of the well known Ising model,
\begin{equation}
H=-K\sum\limits_{<i,j>}S_iS_j,
\end{equation}
where $K=J/k_BT$.
The term $n^{n_c}$ is  factor of cluster weight, and $n_c$ is the number of Ising clusters in 
the configurations and $n$ is a real number.
The clusters are formulated by the connected spins with same directions.

The exploration of the CWI model will help to understand the loop model. By doing similar work like the low temperature expansion~\cite{Low-temperature expasion},
the above model can be transformed into the 
loop model with the relation $x=\exp(-2K)$ between 
the parameters $x$ and $K$. 
The study of the CWI model will helps to understand $n$-component face cubic model~\cite{ding,cubicguowenan,ding13}, whose partition is 
which can transformed into a loop model.

\section{Algorithm and observables}
\label{sec:algorithm}

The algorithm to simulate this model is as follows:
\begin{enumerate}
\item Initialize randomly assigned configuration.
\item
Construct the  Ising clusters: for a pair neighborhood sites  $i$ and $j$ , if $S_i=S_j$,
then absorb site $j$ into to the cluster. 
\item
Assign each Ising cluster with a green color(active) with a probability of $1/n$,
but for a cluster with a red color(inactive) with a probability of  $1-1/n$. 
\item
Construct the Swendsen-Wang clusters: for the site $i$,  add its neighborhood site $j$ into the cluster according the  rules as follows:
 \begin{enumerate}
 \item
No matter what the statuses of  the spins on the  sites $i$ and $j$ are, the only consideration is the color assigned on the sites.
If one site in the site $i$ and $j$ is in red,
then absorb the site $j$ into the cluster absolutely.
\item
If both sites $i$ and $j$ are in green,
then  absorb site $j$ into to the cluster with a probability of $p=1-e^{-2K}$ if $S_i=S_j$.
\end{enumerate}
\item
Flip the clusters with a probability $1/2$.

\end{enumerate}

This algorithm  is precisely  introduced in Ref~\cite{cluster simulation}. In this paper, we only apply it to non-planar graphs. 
 Meanwhile, it would be worth mentioning that there is no particular reason to use Swendsen-Wang algorithm for the Ising updates on the active subgraph in step 4. Actually, ``any" valid Ising Monte Carlo method would suffice, such as worm algorithm ~\cite{worma}, or Sweeny algorithm~\cite{sweeny}, or dynamic connectivity checking algorithm ~\cite{w_e}.

Considered that we assign the clusters by red color (inactive) 
with a probability of $1-1/n$, and hence the  algorithm  we used only works for $n\geq1$ even though the CWI model is well defined for any $n>0$.

With the help of  Monte Carlo algorithm  introduced before, the sampled observables  include the magnetization $m$, the magnetic susceptibility $\chi$, the specific
 heat $C_V$ and the Binder ratio $Q$, which are defined as follows
\begin{eqnarray}
m&=& \mathcal{\langle|M|\rangle},\\
Q&=&{\langle \mathcal{M}^2 \rangle}^2/{\langle\mathcal{M}^4 \rangle},\\
\chi&=&\frac {L^3} {k_BT}[{\langle \mathcal{M}^2 \rangle}-{\langle \mathcal{M} \rangle}^2],\\
C_V&=&\frac 1 {{k_B}T^2}[{\langle E^2 \rangle}-{\langle E \rangle}^2],
\end{eqnarray}
with ${M}$ defined as
\begin{equation}
\mathcal{M}={\sum_i S_i}/{L^3}.
\end{equation}

The previous physical quantities have their scaling behavior as a function of the system size $L$ and the thermodynamic temperature $T$:
\begin{equation}
\begin{aligned}
m=&L^{y_m-d}[m_0+a_1(T-T_c)L^{y_t}+a_2(T-T_c)^2L^{2{y_t}}\\
&+\cdots+b_1L^{y_1}+b_2L^{y_2}+\cdots],
\end{aligned}
\label{m}
\end{equation}
\begin{equation}
\begin{aligned}
Q=&Q_0+e_1(T-T_c)L^{y_t}+e_2(T-T_c)^2L^{2{y_t}}\\
&+\cdots+f_1L^{y_1}+f_2L^{y_2}+\cdots,
\end{aligned}
\label{q}
\end{equation}
where ${T_c}$ is the critical temperature, ${y_t}$ is the thermal exponent, ${y_m}$ is the magnetic exponent, $d$ is the
space dimension and ${y_1}$, ${y_2}$,$\cdots$, are negative correction-to-scaling exponents.
 The expansion coefficients $a_i$, $b_i$, $e_i$, $f_i$, ($i$ = 1, 2,$\cdots$) emerging in the two scaling functions, in general, are different.

The fitting function in Eq.~(\ref{m}) describes how  $m$ depends on the expansion coefficients, and
at the critical points, the function is reduced to
\begin{equation}
m=L^{y_m-d}(m_0+b_1L^{y_1}+b_2L^{y_2}+\cdots),
\label{criticalm}
\end{equation}
which will be used to determine the exponent ${y_m}$.

\section{Results}
\label{sec:results}

\begin{figure}[b]
\includegraphics[scale=0.33,angle=270]{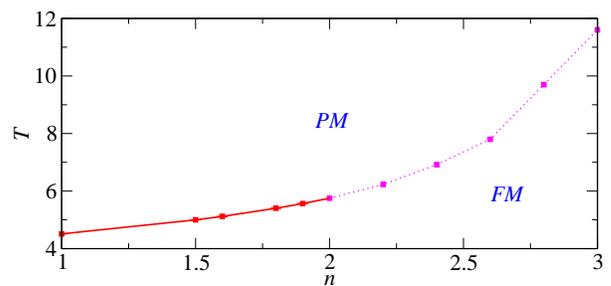}
\vskip -3cm
\caption{(Color online) The global phase diagram in the plane $T_c$ .vs. $n$, containing FM and PM phase. The dashed (solid)  lines denote first (second order) transition.}
\label{fig2}
\end{figure}
Firstly, by the algorithm describing in Sec.~\ref{sec:algorithm}, we perform a Monte Carlo
simulation of the CWI model on the 3D lattice.
The first $10^5 -10^6$ MC steps of simulation is performed in order to let the system reaches 
equilibrium states. 
Then ${10}^5$ samples in each thread (totally 100 threads) are taken to calculate each quantity for
 the system size $16 {\leq} L {\leq} 144$. The estimated auto correlation time $\tau_{int}$ is
 about $L^{0.45(3)}$ explained in Sec.~\ref{sec:atf}. Therefore there are enough independent samples in the total $10^7$ samples. 
To obtain ${T_c}$ and ${y_t}$, we perform a finite-size
scaling analysis of {$Q$} for various system sizes near ${T_c}$.
At {$T_c$},  {$y_m$} is calculated.

\subsection{Global phase diagram and exponents}

To verify our method and results, we first simulate the model with $n=1$ on the 3D lattice equivalent to the 3D Ising model,
  whose critical point is known at {$T_c=4.5110(3)$}~\cite{2}.
Our result $T_c$=4.5115(1) from the Binder ratio $Q$ according to Eq.~(\ref{q}) is consistent with  results in Ref.~\cite{2}.
Apart from the critical points,
$y_t$ and $y_m$ are also very consistent with the results in Ref.~\cite{On3d}.   We obtain $y_t=1.584(4)$ while $y_t$ takes
 value of $1.588(2)$ in Ref.~\cite{On3d}.

The numerical exponents ${y_t}$ and  ${y_m}$ at  the critical points are listed in Table~\ref{table1} for different values of $n$.
The numerical estimation of  $y_t$ and $y_m$ are  different from with values in Ref~\cite{On3d} for $n$ = 1.5.
\begin{table}[t]
 \caption{Comparison of the numerical exponents with those in Ref \cite{On3d}.
 The critical temperature ${T_c}$, the thermal exponent ${y_t}$ and the magnetic exponent ${y_m}$
  for different values of $n$. The estimated errors in the last decimal place are shown between parentheses.}
 \begin{center}
\begin{tabular}{clllllccc}
  \hline
  \hline
~$n$ ~&~ ~~~$T_c$~ & ~~~$y_t$&$y_t${$\rightarrow$}Ref~\cite{On3d}& ~~~$y_m$&$y_m${$\rightarrow$}Ref~\cite{On3d}\\ \hline
1.0&4.5115(1)&1.584(4)&~~1.588(2)&2.487(1)&~~~2.483(3)\\
1.5&4.99912(5)&1.639(6)&~~1.538(4)&2.400(5)&~~~2.482(3)\\
1.6&5.12147(5)&1.71(5)&~~~~~~-&2.34(2)&~~~~~~~-\\
2.0&5.7514(2)&2.95(5)&~~1.488(3)&~~~~~~-&~~~~~~~-\\
  \hline
  \hline
\end{tabular}
\label{table1}
\end{center}
\end{table}

Fig.~\ref{fig2} shows the global phase diagram containing  paramagnetic (PM) phase and ferromagnetic (FM) phase, where the dashed line denotes the first order transition in the range $n_c\leq n<3$ and the solid line represents the second order transition in the range $1<n<n_c$, where $n_c\approx 2$.

To summary, for $n>1$ on the 3D lattice, the universalities of the CWI model and the O($n$) loop model
are  different.


\subsection{$n=1,1.5$, detailed analysis}
\begin{figure}[htpb]
\centering
\includegraphics[angle=270,scale=0.3]{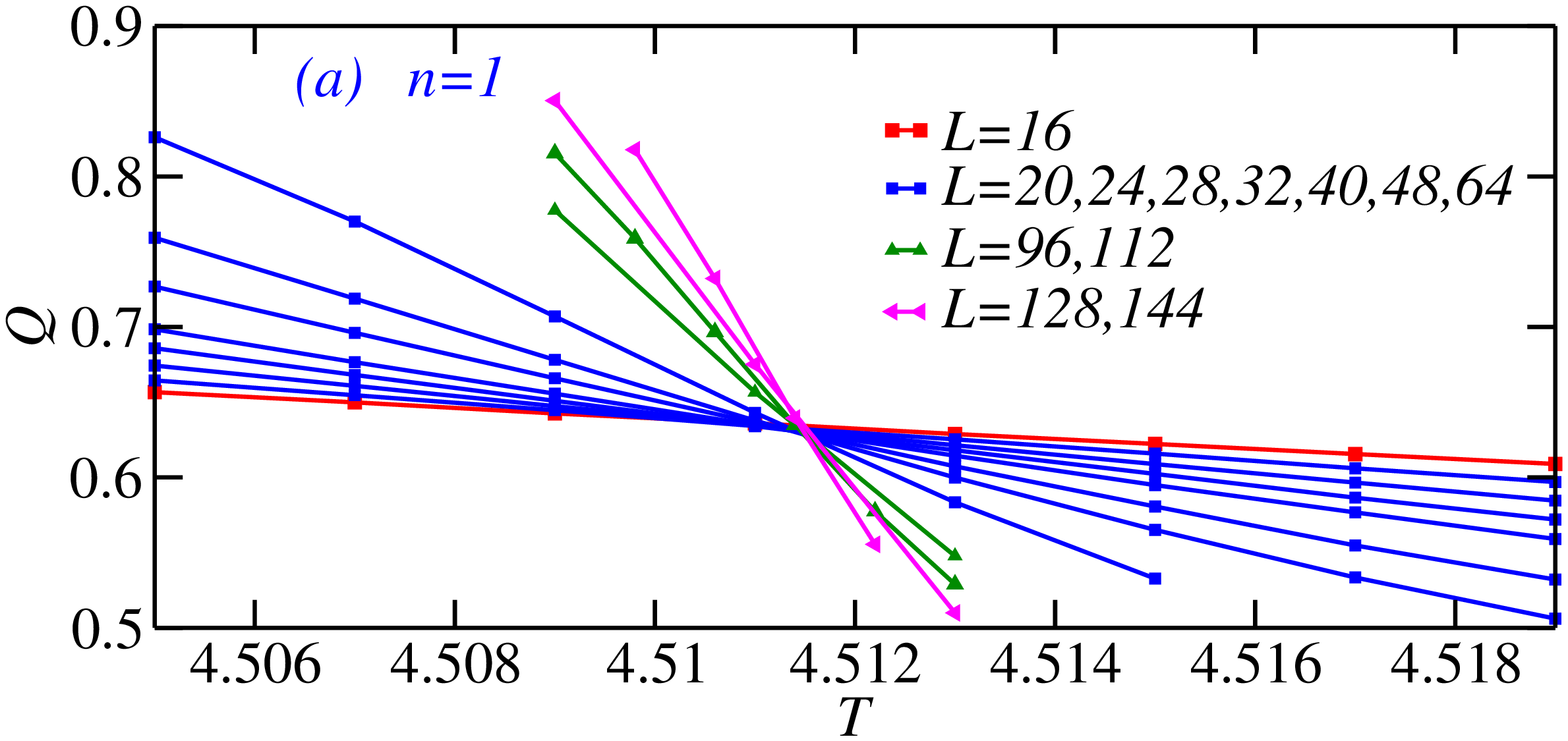}
\vskip -3cm
\includegraphics[angle=270,scale=0.3]{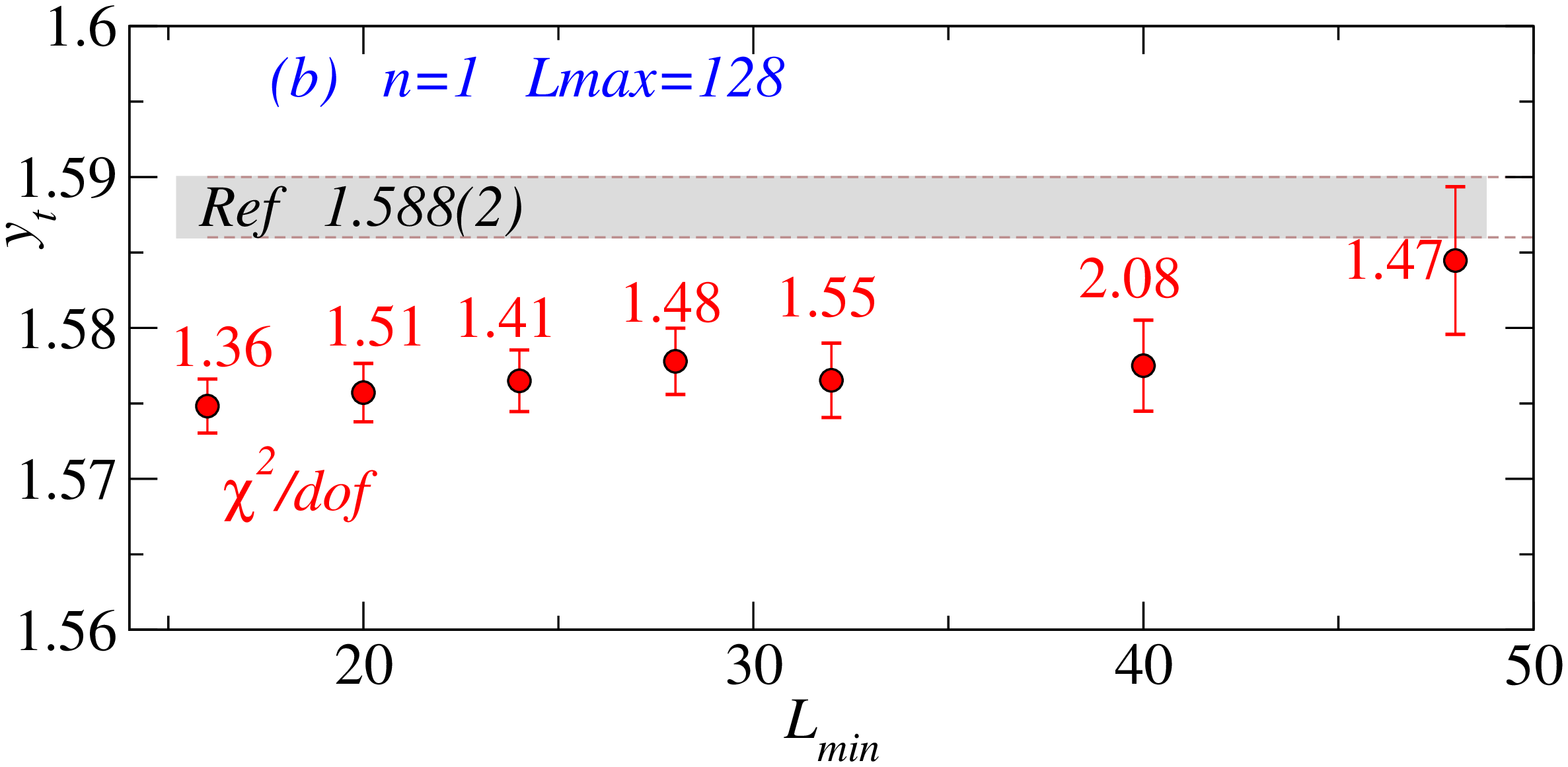}
\vskip -2.8cm
\includegraphics[angle=270,scale=0.3]{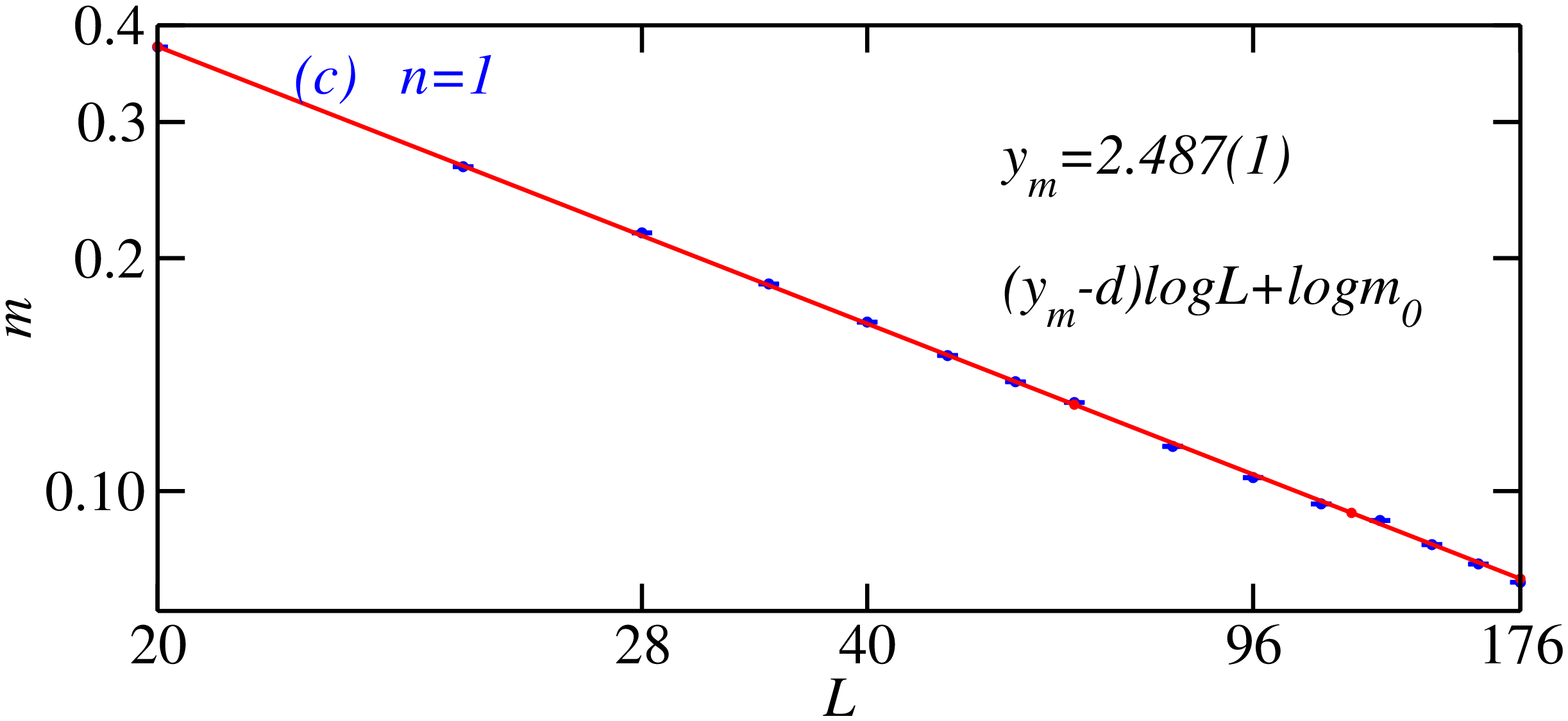}
\vskip -3cm
\caption{(Color online) (a) Binder ratio $Q$ .vs. $T$ at $n = 1$  with different sizes. The critical point is ${T_c} =4.5115(1)$, and $Q_0 = 0.62(4)$. (b) $y_{t}$ .vs. $L_{min}$. The values of  $\chi^2/dof$ are labeled.  (c)The log-log plot of the magnetization $m$ .vs.  $L$.}
\label{n1qt}
\end{figure}
For $n=1$,~the CWI model  is reduced to the pure Ising model, which has been simulated in a higher precision by Ref~\cite{dplandau}.
It provides a good reference to check  our result.
To perform Levenberg-Marquardt(LM)  least-squares fit~\cite{41},  the weighted  distance between data
points and fitting function
is defined as 
$\Delta _ { i } = Q ( T _ { i };  \{ a _ { n } \} ) - Q _ { i }$,
where ${a_n}$ is the parameter to be fitted including the exponents $y_t$, $y_1$, $y_2$, the coefficients $e_1$, $e_2$, $f_1$, $f_2$ and 
other quantities $Q_0$ and $T_c$. 
In practical,  the  error, i.e., standard
deviation $\sigma _ { i }$  of the data points $Q_i$ is divided aiming  to minimise the quadratic distance,
\begin{equation}
\chi ^ { 2 } = \sum _ { i = 1 } ^ { N } \frac { \Delta _ { i } ^ { 2 } } { \sigma _ { i } ^ { 2 } } = \sum _ { i = 1 } ^ { N } \frac { [ Q ( T _ { i }; \{ a _ { n } \} ) - Q _ { i } ] ^ { 2 } } { \sigma _ { i } ^ { 2 } }.
\end{equation}

Figure~\ref{n1qt}(a) show the lines $Q$ .vs. $T$  in the regimes of $T_c$ in the range $4.506< T < 4.518$ with various system sizes from $L=16-144$. 
Using the data $Q$ .vs. $T$ beginning with different values  of  $L_{min}= 16, \cdots, 48$ and  the fixed  maximum size  $L_{max}=144$,  the critical temperature is obtained as  $T_{c}=4.5115(1)$, which is consistent with a more precise value $1/T_c= 0.221 654 626(5)$ \cite{dplandau}.
 
In Fig.~\ref{n1qt} (b),  the red symbols $y_t$ is
obtained by fitting the terms including one corrected term  $f_1L^{y_1}$.  Increasing $L_{min}$, $y_{t}$ gradually converges to a value  $y_{t}=1.584(4)$  consistent to the known result $y_{t}=1.588(2)$ ~\cite{On3d} within the error bars when  $L_{min} = 48$.
The goodness of fit $\chi^2$ per degree is distributed  in an acceptable range between 1.36 and 2.08.   
In Fig.~\ref{n1qt} (c),  the magnetic exponent $y_m=2.487(1)$ is  consistent with the results $y_{m}=2.483(3)$ in Ref~\cite{On3d} according to Eq.~(\ref{m}).

\begin{figure}[htpb]
\includegraphics[scale=0.3]{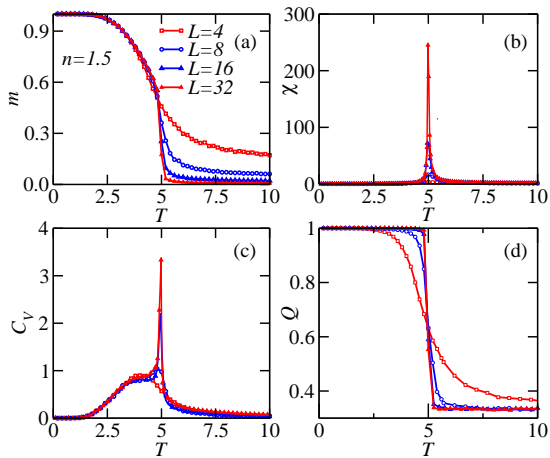}
\caption{(Color online) (a) $m$ (b) ${\chi}$ (c) ${C_V}$ (d) $Q$ .vs. $T$ at $n=1.5$ in the range $0<T<10$, with different sizes $L =4, 8, 16$ and $32$, respectively.}
\label{n1.5mq}
\end{figure}

\begin{figure}[htpb]
\centering
\includegraphics[scale=0.3,angle=270]{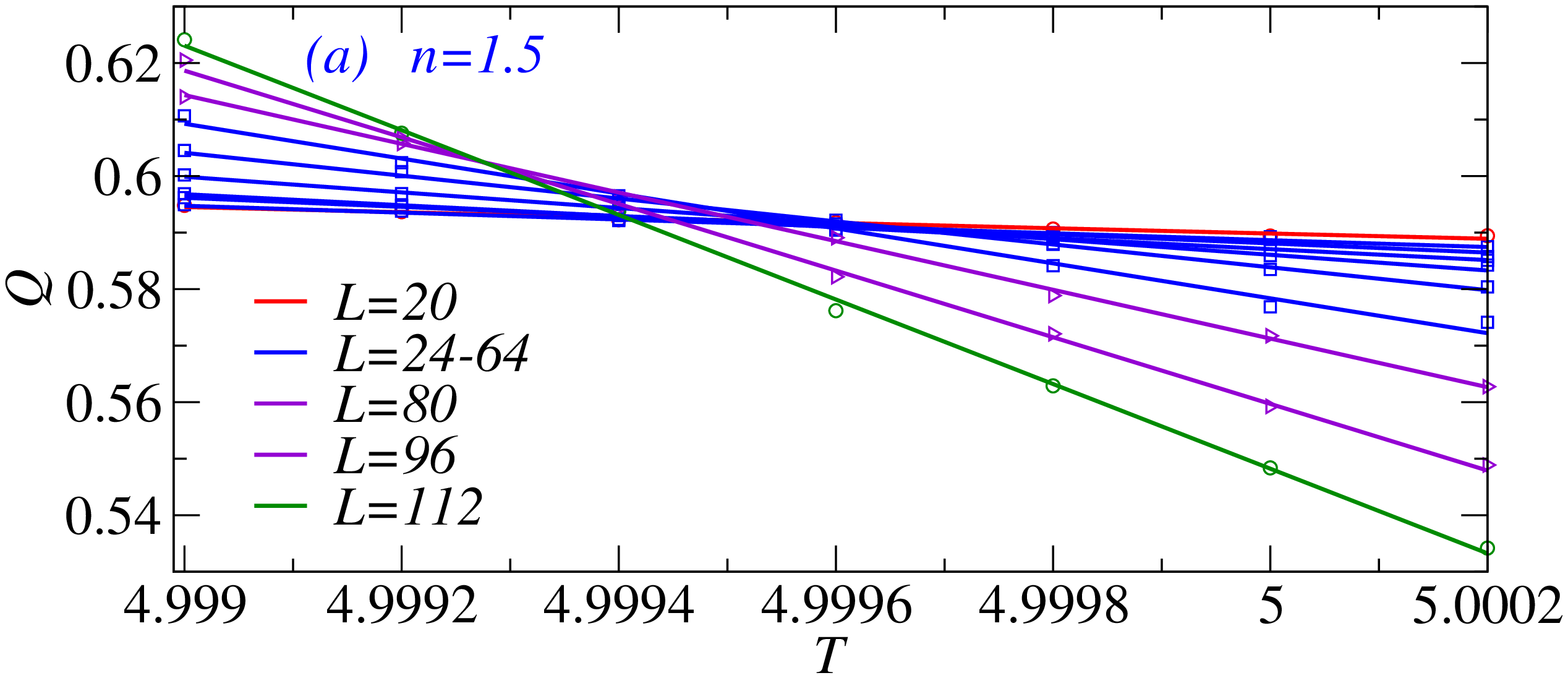}
\vskip -3cm
\includegraphics[scale=0.3,angle=270]{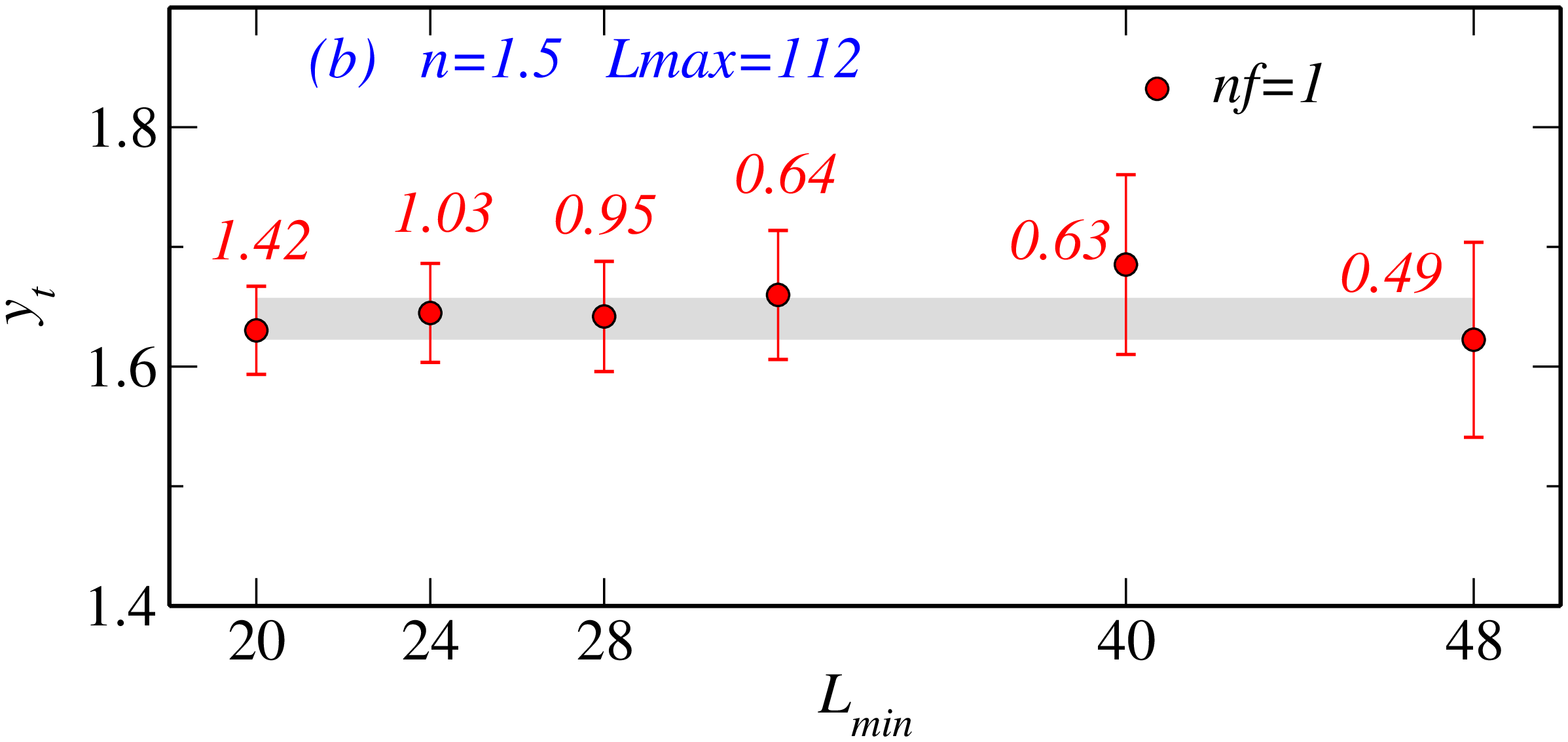}
\vskip -3cm
\includegraphics[angle=270,scale=0.3]{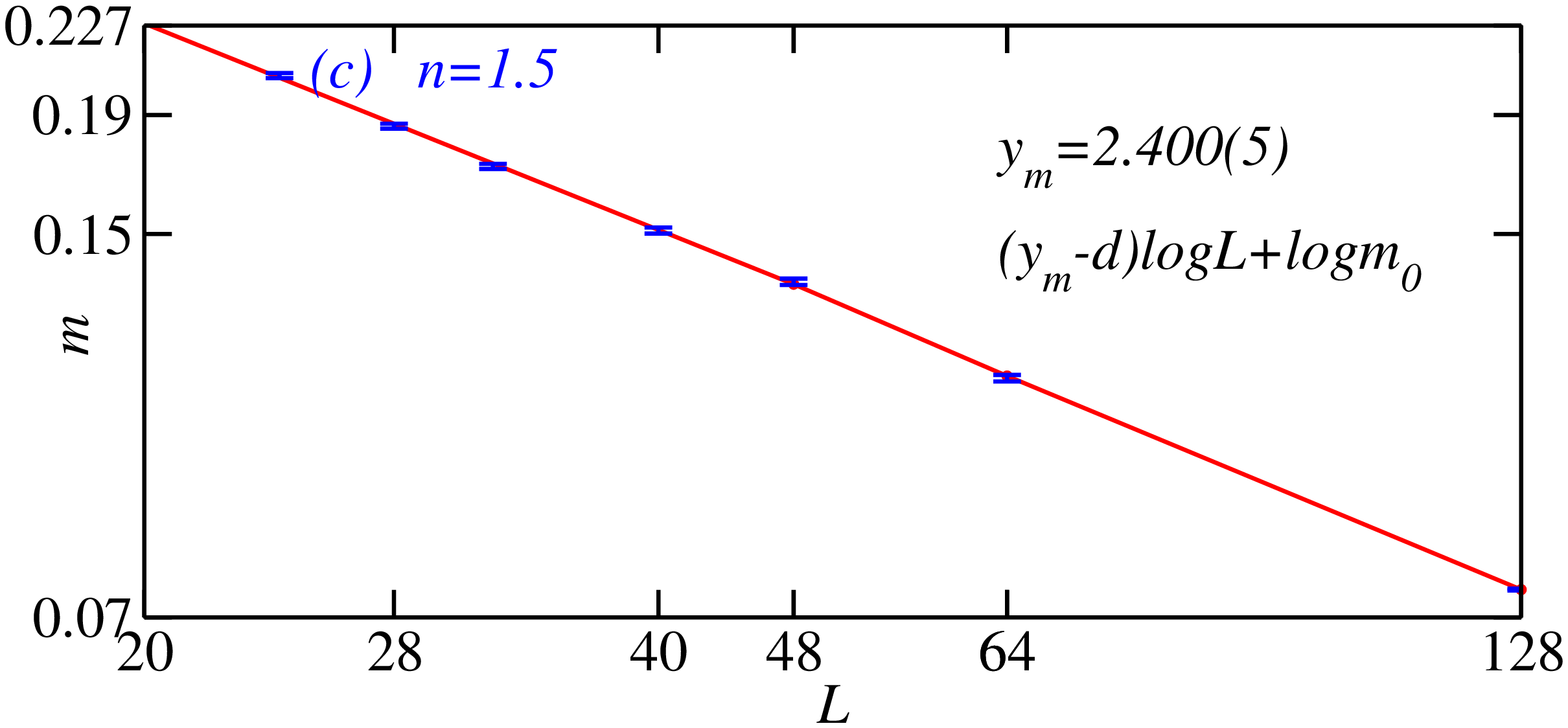}
\vskip -3cm
\caption{(Color online) (a)The Binder ratio $Q$ versus $T$ at $n = 1.5$ with different sizes. (b) $y_{t}$ versus  $L_{min}$. (c) The log-log plot of  $m$ versus $L$.}
\label{QTL}
\end{figure}

For $n=1.5$, to get the the regime of the critical point,  Figure~\ref{n1.5mq} shows the magnetization $m$, the magnetic susceptibility ${\chi}$, the specific heat ${C_V}$ and the Binder ratio $Q$ as a function of $T$  with different system sizes $L=4, 8, 16$ and
 $32$. From the position of peaks and jumps,  $T_c$ is around 5. 

Precise $T_c$ is obtained by  fitting Eq.~(\ref{q}). In  Fig.~\ref{QTL}(a), $Q~.vs.~T$ is  calculated  in a very narrow region   ${4.9990 < T < 5.0002}$  with  many different sizes
$L=4- 144$ and the  the precise critical point is obtained at ${T_c}$=4.99912(5) by the LM algorithm. Correspondingly, the thermal exponent is ${y_t}$=1.639(6). This value is obtain by calcuate the average of $y_t$ through different $L_{min}$, the $\chi^2/dof$ is also shown in  Fig.~\ref{QTL}(b). The obtained $y_t$  is different from    ${y_t}$=1.538(4) in Ref.~\cite{On3d}.
This means that CWI model is not in the same universality as 
the pure loop model ~\cite{On3d}.

Figure~\ref{QTL}(c) shows the log-log plot $m$ versus system size $L$, i.e., $\log (m)=(y_m-d) \log (L)+ \log (m_0 )$ for $n=1.5$ of the CWI model.
The fitted result   $y_m=2.400(5)$ is different from 2.482(3) in the last two significant digits.

\subsection{$n=2, 3$, a first-order phase transition}

We gradually increase  $n$ in the range  $2\leq n<3$. 
 Since first-order transitions are difficult
to study, we first  consider simulating significantly
larger $n$ values, which should be safely in the strongly first-order
regime. The methods used are  ploting the hysteresis and histogram of $m$ and $E$~\cite{1st1,1st2,1st3}, and check whether or not the exponent of $y_t$ equals to $d$.
For $n=2$, histogram and fitting of Eq.~(\ref{q}) are used. The results by different methods check for each other. 

\subsubsection{$n=3$}
\vskip 1cm
\begin{figure}[htpb]
\centering
\includegraphics[scale=0.3,angle=270]{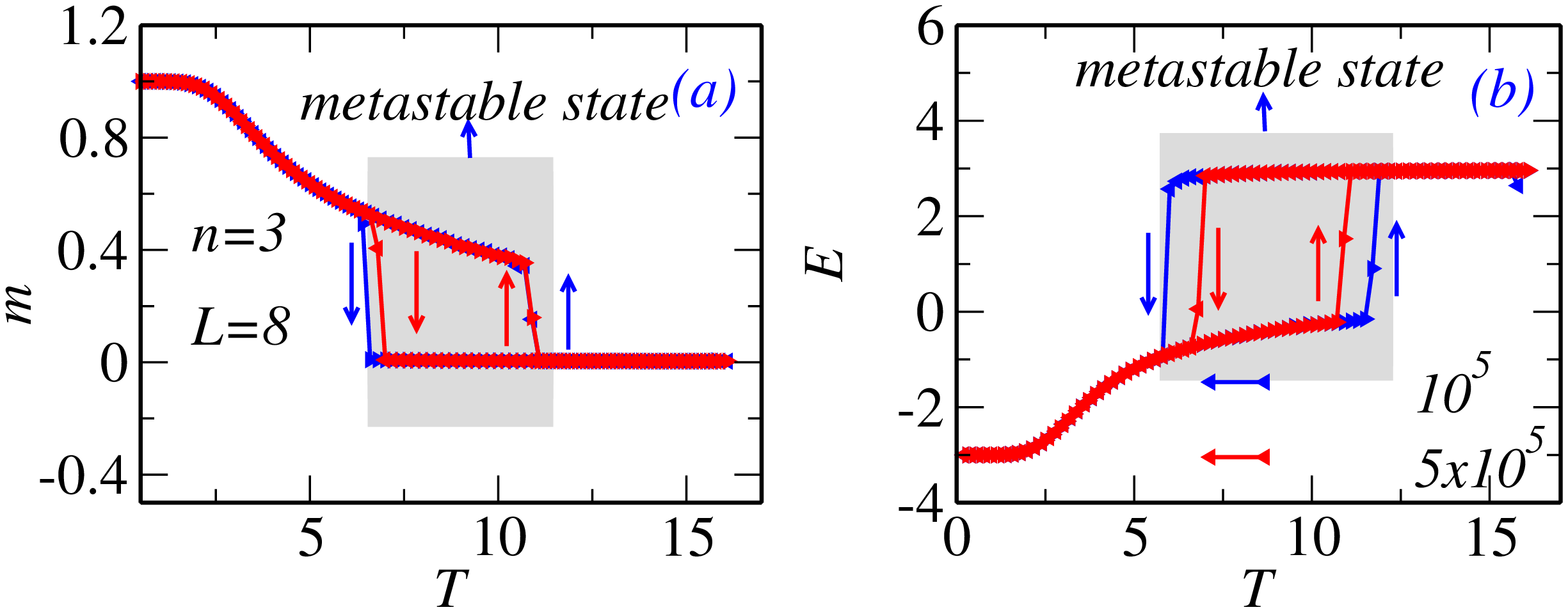}
\vskip -0.3cm
\includegraphics[scale=0.3,angle=270]{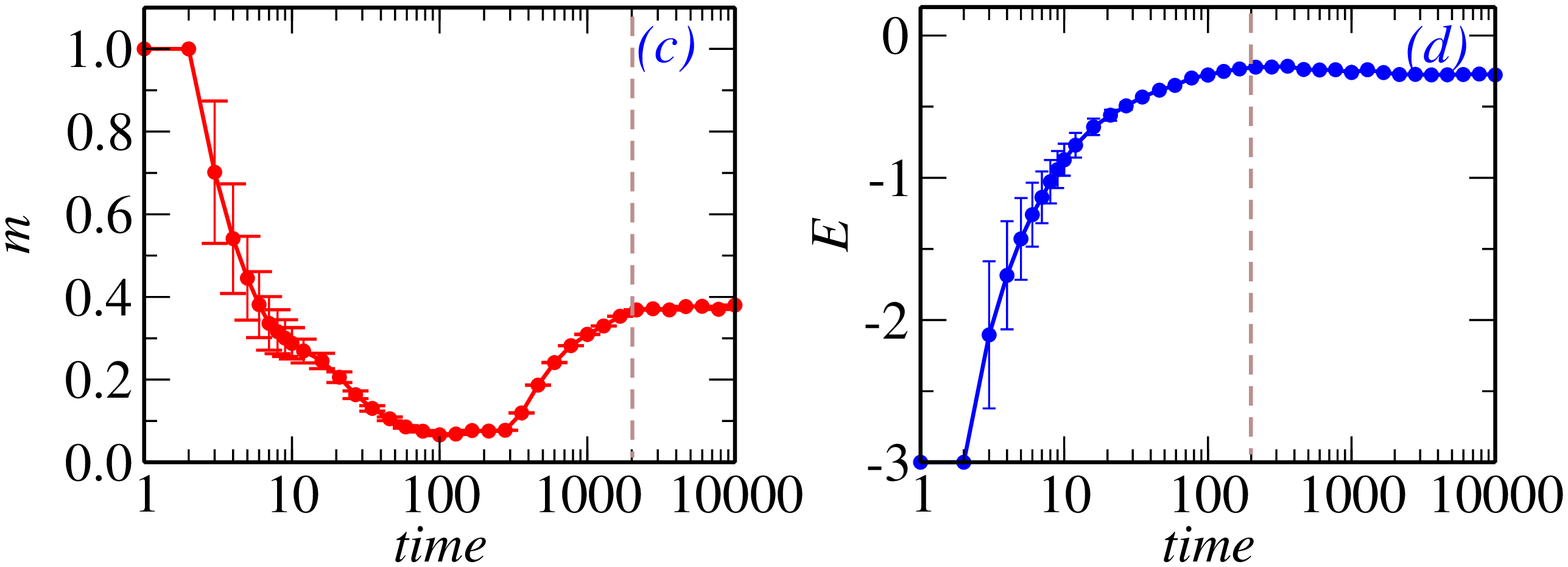}
\vskip -0.3 cm
\includegraphics[scale=0.3,angle=270]{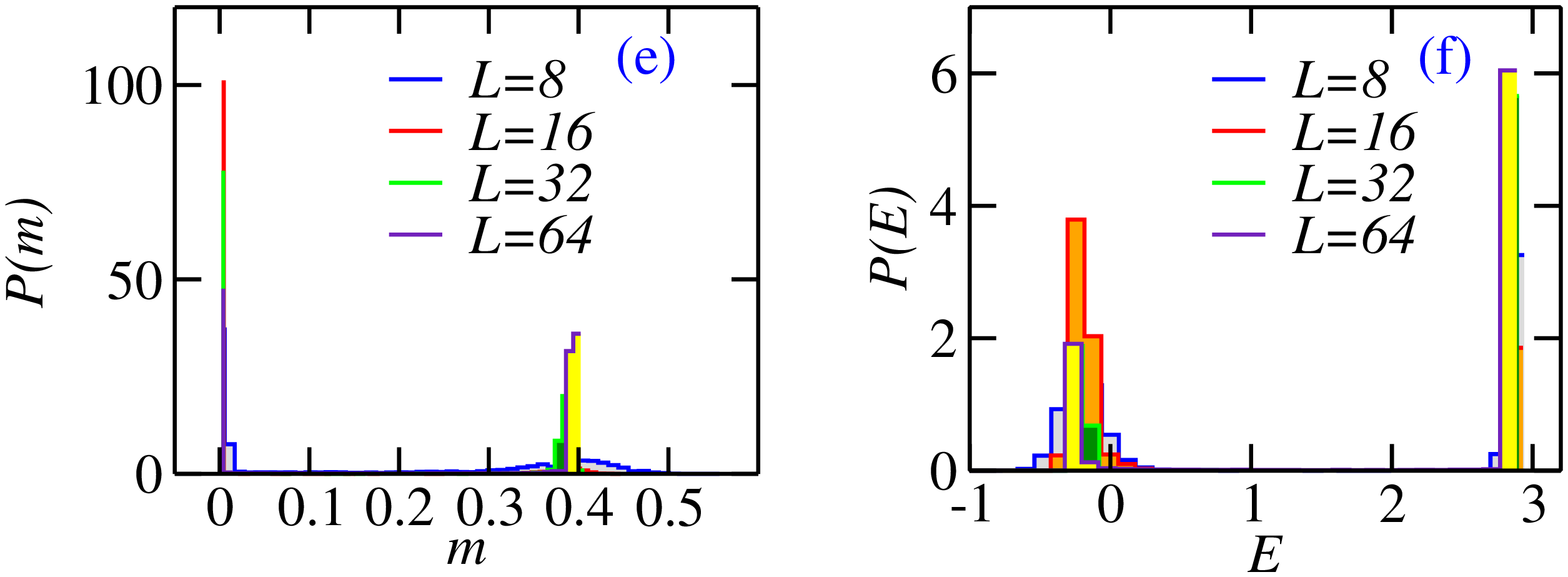}

\caption{(Color online) (a) Hysteresis loop of the quantity $E$ .vs. $T$ (b) the quantity $Q$ .vs. $T$ (c) $m$ (d) $E$ vs. MC time at $T=10$. (e) Double-peak distributions of $m$ at $n=3$, with different sizes $L$ =8, 16, 32 and 64. (f) Double-peak distributions of $E$ at $n=3$, with different sizes $L$ =8, 16, 32 and 64 for $T=$11.8, 11.7, 11.6 and $11.3$.}
\label{n3}
\end{figure}

 To show the signature of a first-order transition for sufficiently large $n=3$, we  draw the 
hysteresis loops of the energy and  magnetization in Fig.~\ref{n3}(a) in the range $0<T<15$.
The expression for the energy $E$ is 
\begin{equation}
E=\frac {[-J\sum_{\langle i,j \rangle} S_iS_j-{n_c}\log (n)]} {L^3},
\end{equation} where $n_c$ is the number of Ising clusters in a spin configuration.
The hysteresis loops have been observed both in classical Baxter-Wu model~\cite{c1} and the site random cluster model~\cite{wss}, as well as quantum systems~\cite{q2,q3,q4},
 which means that there is an obvious first-order phase transition~\cite{book,lan1st}.
 
 Although SW algorithm 
has a global update advantage, but it  still has slow mixing  for a first-order transition regime proved by Ref.~\cite{jst}, which can be used to 
 form a closed hysteresis loop.
Initializing with the temperature  $T=0$,
    we increase  $T$ as well as  sample the energy per site $E$.
   In the simulation of a given value
    of ``$T$'',  we treat the spin configuration  of the  completed simulation, as the (new) initial configuration
    of the simulation of next value of ``$T$''.
    After $T$ exceeds $T_c$ by a small value, the energy per site $E$  jumps to a higher value.  We decrease $T$
     in the same way with regards to the
     initialization of configurations. A closed hysteresis loop shapes when $T$ becomes smaller than $T_c$.
      We repeat these steps in a similar fashion for  $m$, and the loop is shown in Fig.~\ref{n3}(b).


The hysteresis loops are caused by the fact that 
the lifetimes  of the meta-stable states are much longer than the time intervals between  temperature variations~\cite{c1}, and  simulation and the measurements are
taken from meta-stable states,  marked by the gray area.

To confirm above statement, the quantity $\textless m(t)\textgreater=\frac{1}{t}\sum_{l=1}^{t}m(l)$
is also measured, where $m(l)$ is  the observable $m$ observed at time $l$ in the Monte Carlo simulations. $E(t)$ is defined in the same way.
As shown in Fig.~\ref{n3} (c) and (d),  $m(t)$ and $E(t)$ converge  to
0.38013(1)     and -0.27634(2) respectively, which are belong to the values of  one metastable state.

For an infinite system with size $L\rightarrow \infty$, there
will be a discontinuity at $T_c$ of order parameter $m$ ~\cite{book}.
For a finite system,   the probability $p(m)$ is approximated by two Gaussian curves~\cite{lan1st}.
As shown in Figs.~\ref{n2} (e) and (f), there are clearly
 double-peak structures at sizes $L=$ 32 and 64 for the histogram of  $m$ and $E$. The sharp double-peak  at $n=3$ is indicative of sufficiently strong first-order transition.
 
\subsubsection{$n=2$}

\begin{figure}[htpb]
\centering
\includegraphics[scale=0.3,angle=270]{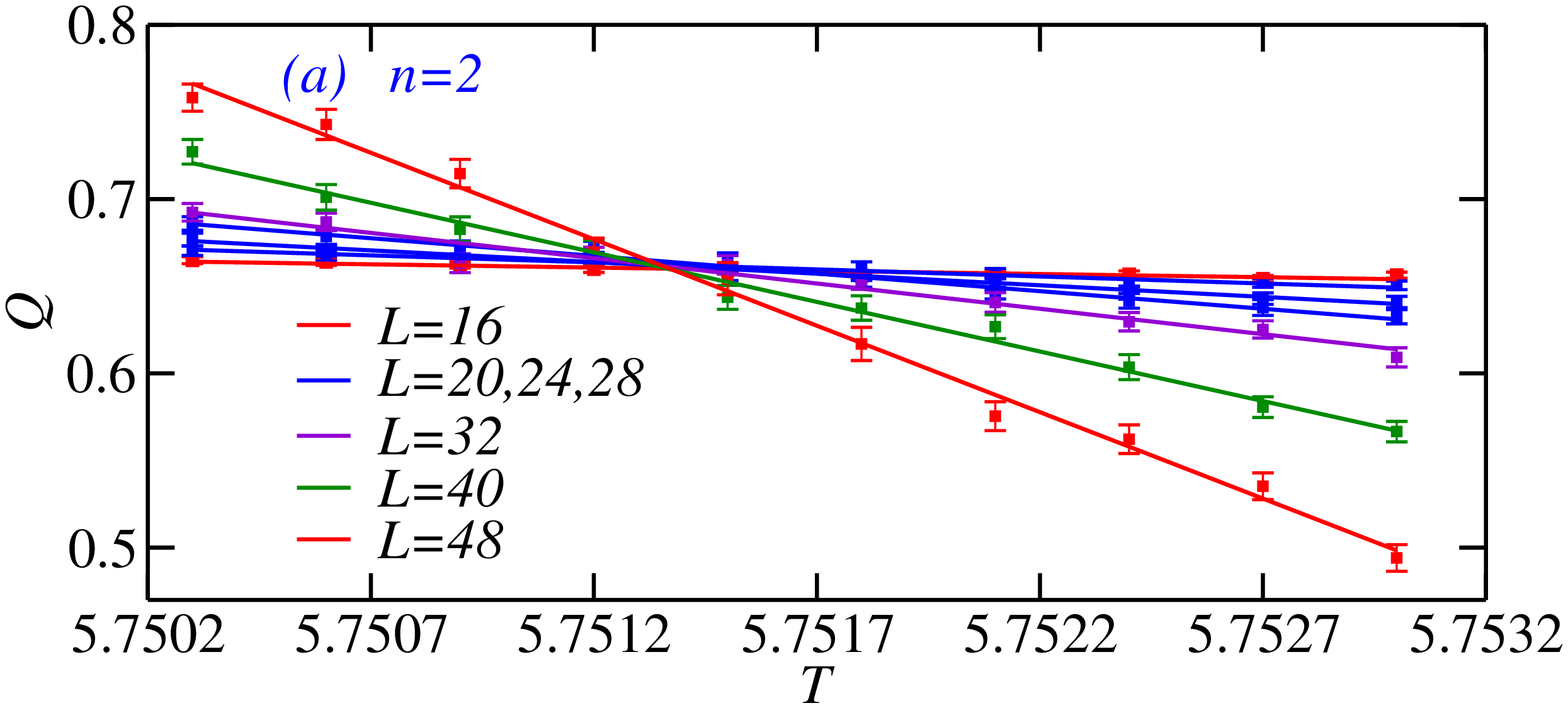}
\vskip -3cm
\includegraphics[scale=0.3,angle=270]{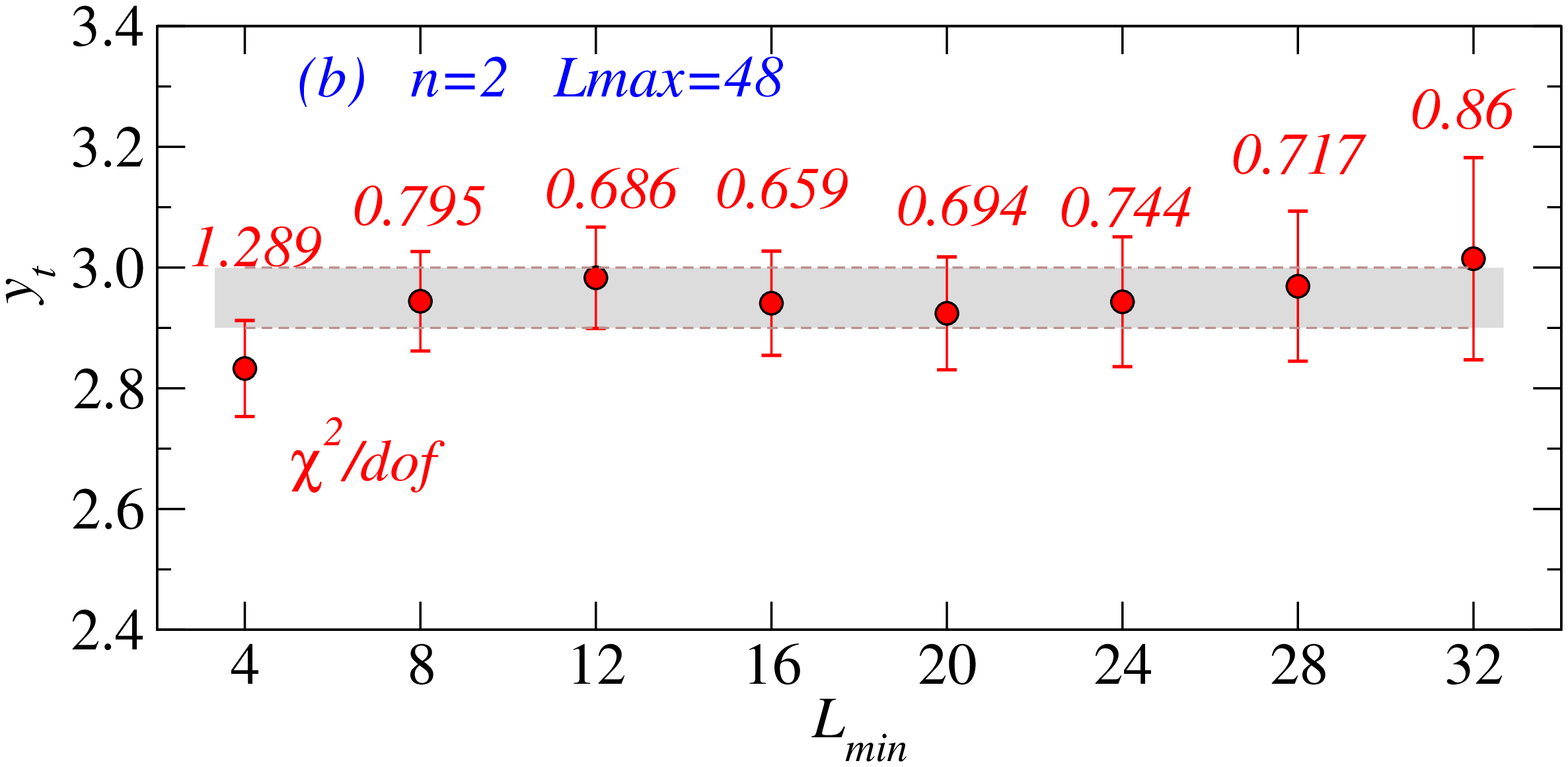}
\vskip -3cm
\includegraphics[scale=0.3,angle=270]{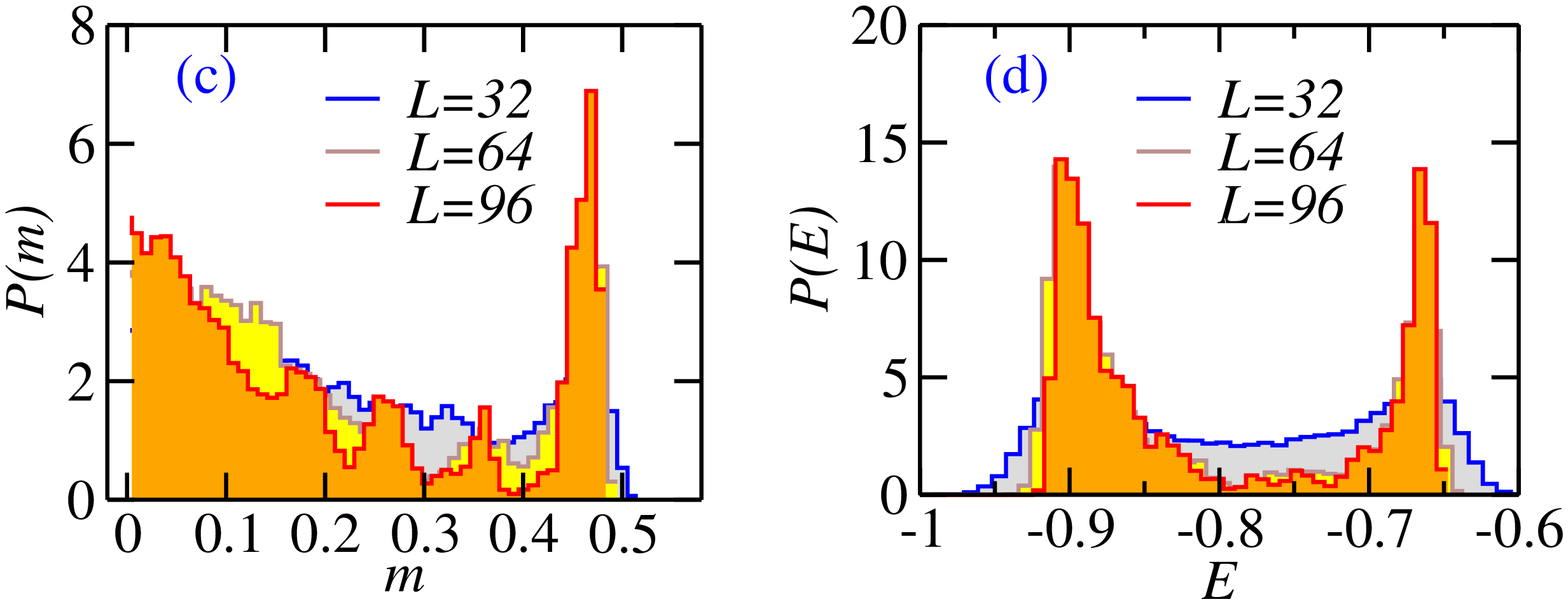}
\caption{(Color online) (a)The Binder ratio $Q$ versus $T$ at $n = 2$ with different sizes. (b) $y_{t}$ versus  $L_{min}$.
(c) Double-peak distributions of $m$ at $n=2$, with different sizes $L$=32, 64, 96. (d) Double-peak distributions of $E$ at $n=2$, with different sizes $L$=32, 64, 96 for $T=$5.755, 5.7518 and 5.7516.}
\label{n2}
\end{figure}

Theoretically, for a first order transition, the fitting of Eq.~(\ref{q}) can not help determine $T_c$\cite{book}. 
However, when  finite system sizes are small   and the temperatures become very close to $T_c$, 
Eq.~(\ref{q}) is used here to determine $y_t$.
Figure~\ref{n2}(a) show the lines $Q$ .vs. $T$  in the regimes of $T_c$ in the range $5.7502< T < 5.7532$ with various system sizes from $L=16-48$. By performing the LM algorithm, 
the values of goodness of fit $\chi^2/dof$ are acceptable and the values are  shown in  Figs~\ref{n2}(b). By sum over $y_t$ with different values of $L_min$, the average $y_t$  is obtained as  2.95(5) indicating the scaling dimension $y_t$ equals to the space dimension $d$, i.e., $y_t=d$. This result is 
consistent with the conclusions in Refs~\cite{1st1,1st2,1st3}.

In  Figs~\ref{n2} (c) and (d), the double-peak distributions of $m$ and  $E$ at $n=2$ are shown, with different sizes $L$=32, 64, 96  for $T=$5.755, 5.7518 and 5.7516.
Increasing the system sizes, the peaks become sharper representing that a first-order transition occurs.

\subsection{Autocorrelation function}
\label{sec:atf}

The algorithm has a little critical slowing down phenomenon. 
This phenomena can be judged by   autocorrelation time $\tau_{int}$,
which can identify the number of 
 MC steps required between two configurations before they can
be considered statistically independent~\cite{aws}.

For the quantity absolute magnetism $M=|\mathcal{M}|$, the integrated autocorrelation function
$A_{M}(t)$ is defined as:
\begin{equation}
A_{M}(t)=\frac{\langle M_{k}M_{k+t}\rangle-\langle M_{k}\rangle^2}{\langle M_{k}^2\rangle-\langle M_{k}\rangle^2}\end{equation}
and the integrated accucorrelation time $\tau_{int}$ is defined as 
\begin{equation}
\tau_{int}=\frac{1}{2}+\sum_{t=0}^\infty A_{M}(t)\end{equation}

\begin{figure}[htpb]
\vskip 1 cm
\centering
\includegraphics[angle=270,scale=0.3]{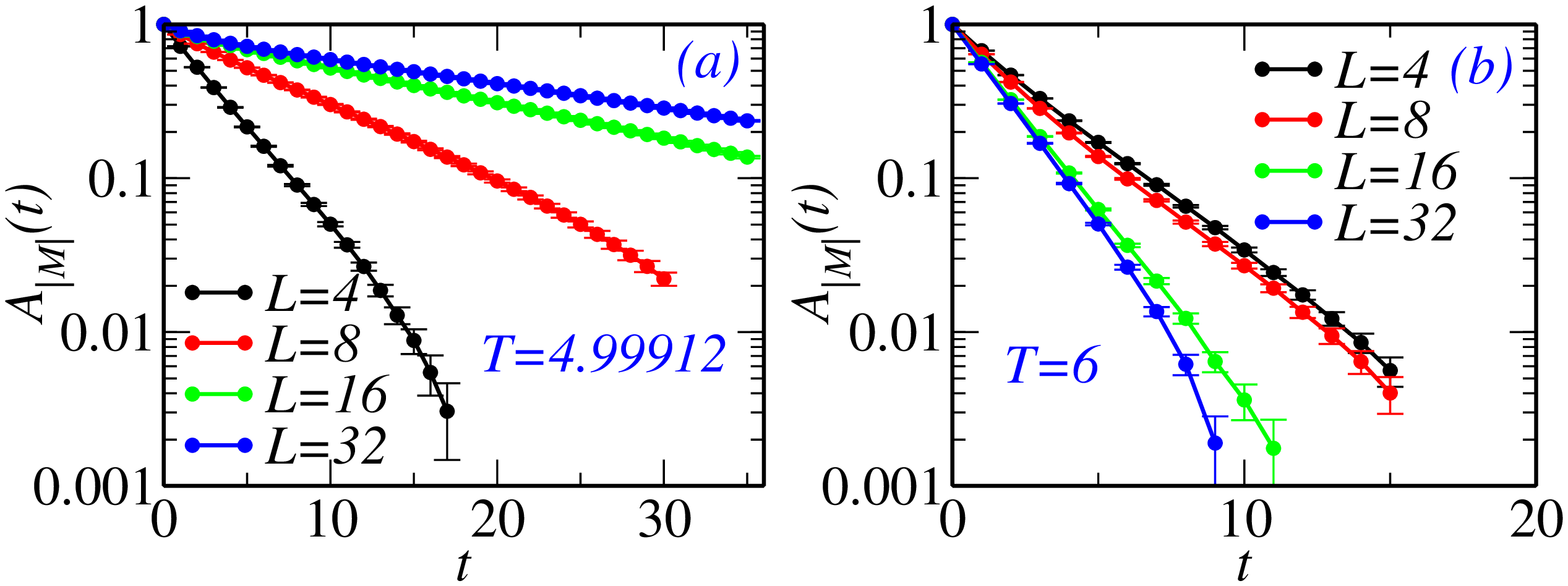}
\includegraphics[scale=0.32,angle=270]{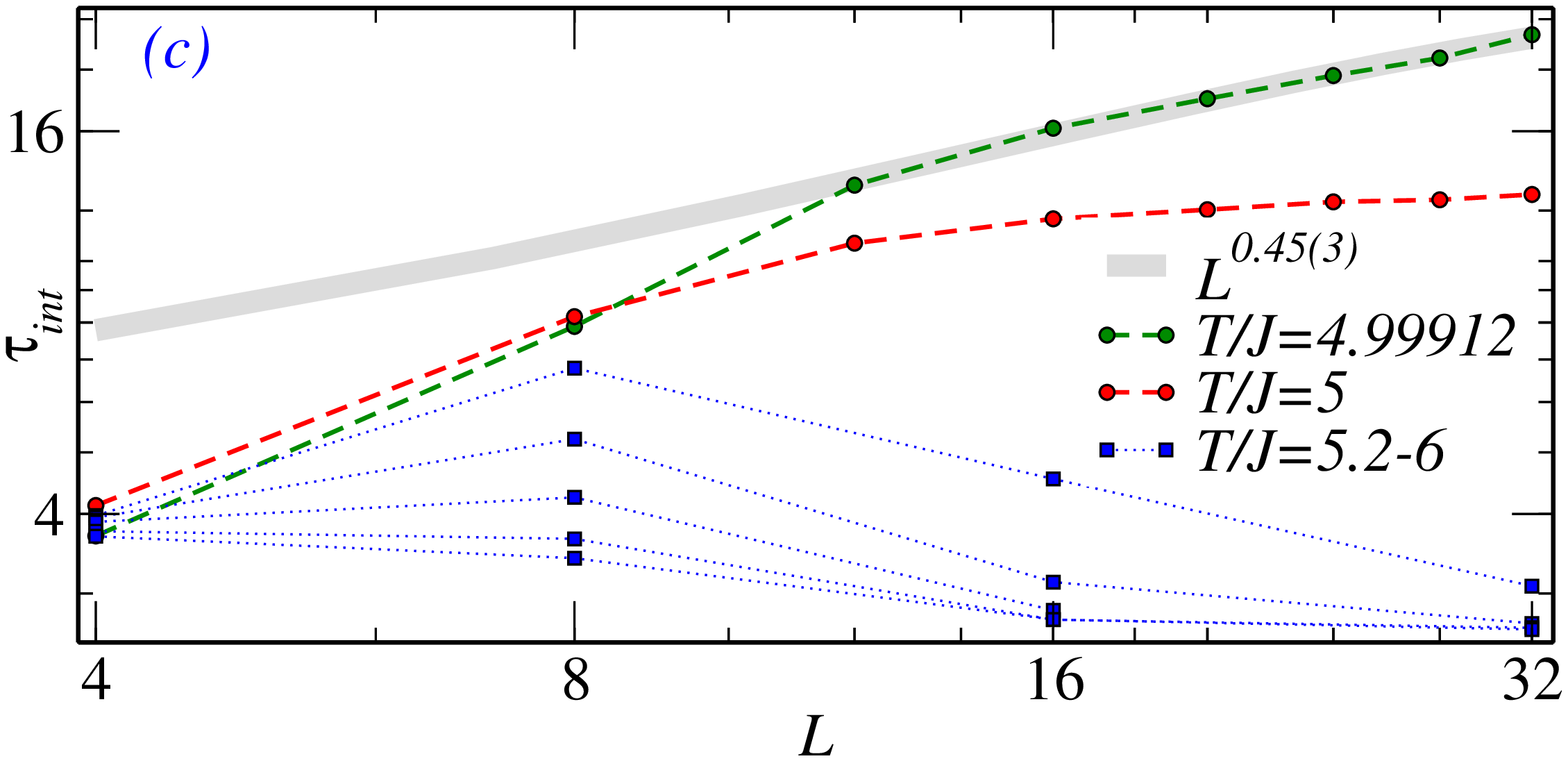}
\vskip -3cm
\caption{(Color online) 
$A_{M}(t)$ versus $t$ for $n=1.5$ CWI model with  different sizes at (a) $T=T_c$ (b)  $T=6>T_c$. (c) $\tau_{int}$ versus $L$ at different temperatures.
The exponent $z$ is fitted to be  $z=0.45(3)$.}
\label{n1.5zxg}
\end{figure}

In Figs.~\ref{n1.5zxg}(a) and (b), the autocorrelation function $A_{m}(t)$ for the $M$  decays almost purely exponentially on MC time (a linear decay on the linear-log scale).
Close to $T_{c}=0.499912$, $A_{M}(t)$  grows  with $L$ while it decreases with $L$ when the temperature 
deviates away from  $T_{c}$.

As shown in Fig.~\ref{n1.5zxg}(c),  the integrated autocorrelation time $\tau_{int}$  behaviors in the way like
$L^{z}$ at $T=T_c$  where the dynamical exponent
$z=0.45(3)$ and the error bar in the parentheses is the systematic error due to corrections to scaling~\cite{Sokal}. This value of $z$
is consistent (within error bar) with the result   0.443±0.005±0.030 of a ``susceptibility-like'' observable, or 0.459±0.005±0.025 of a ``energy like'' observable  from Ref.~\cite{Sokal} studying the three dimensional Ising model by the SW algorithm~\cite{SW}. 
The algorithm we used for $n=1.5$ is as efficient as the SW algorithm.

\subsection{Error bar analysis} 
The programs are run in 100 threads (bins), and in each thread (bin) different seed of random number generator is used.  The first $10^5-10^6$  MC steps of simulations are run without measuring any quantities  allowing the systems to reach the stage of  equilibrium.   $10^7$ times of sampling are performed in the equilibrium states and the mean values of the quantities are collected from each bin. 

For example, the quantity $m$ is  average from many  bins according to $m=\frac{1}{nbin}\sum_{b=1}^{nbin}\overline{m}_{b}$, where $\overline{m}_{b}$, $b=1, \cdots, nbin$   are computed over each bin.
The error bar $\sigma$ is calculated according to \begin{equation}
\sigma=\sqrt{ \frac{1}{nbin(nbin-1)}\sum_{b=1}^{nbin}(\overline{m}_{b}-m)^2}
\end{equation}
The quoted error bar corresponds to one
standard deviation (i.e., confidence level $\approx 68\%$).

The error bars of the fitted exponents are estimated by the diagonal  elements of the  covariance matrix $[C]=[\alpha]^{-1}$, where $\alpha$ is  defined by~\cite{41}

\begin{equation}\alpha_{kl}=\sum_{i=1}^{N}\frac{1}{\sigma_{i}^{2}}\left[\frac{\partial Q ( T _ { i }; \{ a _ { n } \} )}{\partial a_{k}}\frac{\partial Q ( T _ { i }; \{ a _ { n } \} )}{\partial a_{l}}\right]\end{equation}

\section{ Discussion and conclusion}
\label{sec:conclusion}

It should be noted that,  loop model can be obtained as a high-temperature expansion of various cubic models, such as both face and corner cubic model~
\cite{ncubic}, in which the spins point to
the corners of an $n$-dimensional hypercube, but it also arises as a high temperature expansion of the O($n$) vector spin model in certain settings. The descriptor cubic or O($n$) refers to a symmetry of the spin Hamiltonian, and has no immediate interpretation in the loop language. Furthermore, the face cubic and corner cubic models can both be related to this same loop model but can have entirely different phase transitions~\cite{ncubic}.

In conclusion, we have proposed a cluster weight Ising (CWI) model,
composed of an Ising model with an additional
 cluster weight in the partition function with respect to the traditional Ising model. In order to simulate the CWI model, we  apply an
efficient cluster algorithm  by combining  the color-assignation and the Swendsen-Wang method.
The algorithm has almost the   same efficiency as the Swendsen-Wang method, i.e., 
the dynamical exponent for the absolute magnetization  $z=0.45(3)$ at $n=1.5$ is   consistent with that of  the traditional Swendsen-Wang method.

Second order transitions
 emerges 
 with $1\leq n< n_c$ and first order transitions 
occur when  $n \geq n_c$ ($n_c \approx 2$) of the CWI model on the 3D lattices, and the universalities of our CWI model and the loop model~\cite{On3d}
are completely different.
The  first-order transition is verified by the signatures  of  hysteresis,  double-peak structure of histograms for the order parameters, and the value of the critical exponent $y_t=d$.
Our results can be helpful in the understanding of traditional statistical models.

\section*{Acknowledgments}
We thank Prof. Youjin Deng for his  discussions  and the valuable suggestions from the referees, also thank T. C. Scott for helping  prepare this manuscript.
C. Ding is supported by the NSFC under Grant No. 11205005 and Anhui Provincial Natural
 Science Foundation under Grant No. 1508085QA05.
 W. Zhang is supported by the
 open project KQI201 from 
 Key Laboratory of Quantum Information, University of Science and Technology of China, Chinese Academy of Sciences.

 




\end{document}